\documentclass[sigconf, nonacm]{acmart}

\setcopyright{none}
\renewcommand\footnotetextcopyrightpermission[1]{}

\copyrightyear{2026}
\acmYear{2026}
\setcopyright{acmlicensed}

\renewcommand\footnotetextcopyrightpermission[1]{}



\author{Tao Liu}
\affiliation{%
  \institution{Queensland University of Technology}
  \city{Brisbane}
  \state{QLD}
  \country{Australia}
}
\email{t24.liu@qut.edu.au}

\author{Gowri Ramachandran}
\affiliation{%
  \institution{Queensland University of Technology}
  \city{Brisbane}
  \state{QLD}
  \country{Australia}
}
\email{g.ramachandran@qut.edu.au}

\author{Raja Jurdak}
\affiliation{%
  \institution{Queensland University of Technology}
  \city{Brisbane}
  \state{QLD}
  \country{Australia}
}
\email{r.jurdak@qut.edu.au}

\graphicspath{{figures/results/}}

\usepackage{booktabs}
\usepackage{amsmath}
\usepackage{graphicx}
\usepackage{caption}
\usepackage{subcaption}
\usepackage{array}
\usepackage{tabularx}
\usepackage{multirow}
\usepackage{makecell}
\usepackage{enumitem}

\newcolumntype{L}{>{\raggedright\arraybackslash}X}
\newcolumntype{C}{>{\centering\arraybackslash}X}
\newcommand{\rowstyle}[1]{%
  \gdef\rowstyle@{#1}#1\ignorespaces}
\newcolumntype{R}[1]{>{\rowstyle@}#1}

\usepackage{pifont}

\begin{document}

\title{On the Energy Cost of Post-Quantum Key Establishment in Wireless Low-Power Personal Area Networks}

\begin{abstract}

Post-Quantum Cryptography (PQC) creates payloads that strain the timing and energy budgets of Personal Area Networks. In post-quantum key exchange (PQKE), this causes severe fragmentation, prolonged radio activity, and high transmission overhead on low-power wireless devices. Prior work optimizes cryptographic computation but largely ignores communication cost. This paper separates computation and communication costs using Bluetooth Low Energy as a representative platform and validates them on real hardware. Results show communication often dominates PQKE energy, exceeding cryptographic cost. Efficient quantum-resilient pairing therefore requires coordinated protocol configuration and lower-layer optimization. This work provides developers a practical way to reason about PQC energy trade-offs and informs the evolution of PAN standards toward quantum-safe operation.

\end{abstract}

\maketitle

\section{Introduction}

The potential emergence of large-scale quantum computers threatens widely deployed public-key cryptosystems such as RSA and Elliptic Curve Cryptography \cite{nist_sp_1800_38}. In response, standardization efforts led by NIST have produced post-quantum cryptographic (PQC) algorithms for key encapsulation and digital signatures that are believed to resist quantum attacks \cite{NIST_IR_8547_ipd}. These algorithms, however, introduce substantially larger public keys and ciphertexts than pre-quantum schemes. Personal Area Networks (PANs), however, are optimized for strict energy budgets. As a result, integrating PQC into PANs significantly increases communication overhead and stresses physical-layer mechanisms that dominate energy consumption. PQC introduces three fundamental costs: increased cryptographic computation, extensive packet fragmentation, and prolonged radio activity. These costs directly impact energy consumption and latency, making PQC's feasibility a first-order concern for low-power wireless PANs. 

In this paper, we evaluate Post-Quantum key exchange (PQKE) using Bluetooth Low Energy (BLE) as a representative stack. BLE provides a connection-oriented architecture with configurable parameters such as the ATTribute Layer MTU (ATT MTU) and Link Layer (LL) PDU size, enabling controlled and reproducible measurements. We decompose PQKE energy into computation and communication components and validate this decomposition through real hardware measurements, showing how post-quantum payload sizes interact with fragmentation, acknowledgments, and link-layer scheduling. While our evaluation focuses on BLE, the observed communication-energy effects arise from general PAN constraints. 

Our objective is to provide a measurement-driven characterization of the energy cost of PQKE in low-power wireless PANs. By identifying dominant cost components, quantifying their dependence on protocol configuration, and validating results experimentally, this work offers concrete design insights for integrating post-quantum security into PANs, clarifying where optimization efforts are most effective.

The contributions of this work are as follows:

\begin{enumerate}[,itemsep=2pt,topsep=2pt]
\item \textbf{Measurement-driven characterization of PQKE cost.}
We present a detailed empirical study of PQKE on BLE, using real hardware to quantify computation and communication energy across security levels and protocol configurations.

\item \textbf{Identification of dominant protocol-induced energy costs.}
Through theoretical modeling and empirical measurements, we show that communication-side effects can dominate PQKE energy consumption, revealing costs not captured by previous analysis.

\item \textbf{Design guidance and cross-technology applicability.}
Building on the results, we derive guidelines for selecting energy-optimal link parameters under post-quantum security constraints. We further discuss how the modeling principles extend to other low-power personal-area networks to guide future quantum-resilient PAN design.
\end{enumerate}

\section{Background and Related Work} \label{background}

\subsection{Review on PQC Migration}

Large-scale quantum computing threatens widely deployed public-key cryptosystems such as RSA and Elliptic Curve Cryptography (ECC). Shor’s algorithm \cite{shor1994algorithms} enables efficient integer factorization and discrete logarithm computation on quantum computers, rendering these schemes insecure once sufficiently powerful machines become available. The “harvest now, decrypt later” (HNDL) threat model \cite{mosca2018} has increased the urgency of transitioning security protocols toward quantum-resistant alternatives, particularly for data with long confidentiality lifetimes.

In response, the United States National Institute of Standards and Technology (NIST) initiated a multi-round PQC standardization process to identify Key Encapsulation Mechanisms (KEMs) and signature schemes secure against both classical and quantum adversaries \cite{NIST_IR_8547_ipd}. Candidate algorithms underwent extensive cryptanalysis and performance evaluation, but ultimately this transition is challenging for embedded and low-power systems, as PQC schemes incur larger public keys and ciphertexts as well as higher computational costs than pre-quantum counterparts, stressing both communication protocols and energy budgets \cite{khalid2019lattice}.

In 2022, NIST announced the first algorithms selected for standardization. For key establishment, the module-lattice-based ML-KEM (derived from CRYSTALS-Kyber \cite{kyber_paper}) was chosen, while ML-DSA (based on Dilithium \cite{dilithiun_paper}) was selected for digital signatures. NIST also standardized the stateless hash-based scheme SPHINCS+ \cite{bernstein2015sphincs} and later expanded its KEM portfolio with the code-based scheme HQC \cite{HQC_Spec_2025} to diversify hardness assumptions. Table~\ref{tab:nist_pqc} summarizes the standardized PQC algorithms alongside classical ECC baselines. As shown, PQC artifacts are one to two orders of magnitude larger than their ECC counterparts.

\begin{table}[t]
\centering
\caption{NIST-Selected PQC Schemes, with Classical ECDH and ECDSA Included for Payload Size Comparison}
\label{tab:nist_pqc}
\scriptsize
\setlength{\tabcolsep}{4.5pt}
\renewcommand{\arraystretch}{1.15}
\begin{tabularx}{\columnwidth}{l c c c c}
\toprule
\textbf{ } &
\textbf{Type} &
\makecell{\textbf{Public Key}\\\textbf{Size (B)}} &
\makecell{\textbf{Secret Key}\\\textbf{Size (B)}} &
\makecell{\textbf{CT / Sig}\\\textbf{Size (B)}} \\
\midrule

\textbf{ECDH (P-256)} &
KEM &
65 &
32 &
65 \\

\textbf{ML-KEM} &
KEM &
800 / 1{,}184 / 1{,}568 &
1{,}632 / 2{,}400 / 3{,}168 &
768 / 1{,}088 / 1{,}568 \\

\textbf{HQC} &
KEM &
2{,}249 / 4{,}522 / 7{,}245 &
56 / 64 / 72 &
4{,}481 / 9{,}026 / 14{,}485 \\

\midrule

\textbf{ECDSA (P-256)} &
Sig. &
65 &
32 &
64 \\

\textbf{ML-DSA} &
Sig. &
1{,}312 / 1{,}952 / 2{,}592 &
2{,}560 / 4{,}032 / 4{,}896 &
2{,}420 / 3{,}309 / 4{,}627 \\

\textbf{Falcon} &
Sig. &
897 / 1{,}793 &
1{,}281 / 2{,}305 &
666 / 1{,}280 \\

\textbf{SPHINCS+} &
Sig. &
32 / 48 / 64 &
64 / 96 / 128 &
7{,}856--49{,}856 \\

\bottomrule
\end{tabularx}
\end{table}

\subsection{Prior Work on Quantum-Safe IoT}

\subsubsection{Benchmarking PQC for Embedded Systems}

With the standardization of ML–KEM and ML–DSA \cite{nist-fips-203-2024, nist-fips-204-2024}, the research focus has shifted to quantifying cost on real embedded platforms. Existing work primarily reports cycle counts and memory usage, while the impact of large post-quantum payloads on transmission energy remains largely unaddressed.

The pqm4 benchmark project \cite{pqm4} systematizes cycle-level evaluation of PQKEMs and signatures on Arm Cortex-M4, establishing widely used baselines. Subsequent work optimizes lattice-based ciphers through improved arithmetic and data movement, achieving performance gains on constrained MCUs \cite{abdulrahman2022fasterm4, botros2019kyberm4}. While these studies provide realistic computational baselines, they do not capture the energy impact of wireless communication.

Several studies examine PQC integration in embedded and IoT systems but remain largely software-centric. Works such as \cite{hernandez2022ACS} and \cite{maryam2025benchmark} focus on cycle counts and memory footprints, omitting the effects of fragmentation and radio activity that dominate energy consumption in low-power PANs. This gap motivates the need for system-level evaluation that jointly considers all cost components when assessing PQKE feasibility.

\subsubsection{System-level PQC Integration}

System-level studies integrate PQC into network protocols and analyze end-to-end effects. For example, \cite{Mcloughlin2023FullPQDTLS} instantiates a post-quantum DTLS 1.3 stack on IoT hardware, while \cite{GonzalezWiggers2022kemtls} re-architects the handshake using KEM-based authentication. EPQUIC \cite{dong2025epquic} integrates Kyber-based PQKE into QUIC/TLS for embedded platforms, and \cite{malina2024mqtt} implements quantum-resistant security directly within MQTT payloads.

While most system-level work targets Internet-oriented protocols, fewer studies focus on local wireless technologies. For example, \cite{liu2024towards} retrofits BLE encryption with Kyber-512 and examines fragmentation and energy trade-offs, whereas \cite{bozhko2023pqtls} runs TLS over IP-over-BLE and Wi-Fi without modifying the underlying wireless security protocol. In contrast, this work focuses directly on the energy implications of PQKE within native low-power PAN stacks.

\subsubsection{Hardware Acceleration for Quantum-Resilient IoT}

Hardware acceleration offers an orthogonal approach to reducing PQKE cost. Recent work explores PQC accelerators integrated into IoT-class processors and roots of trust, including RISC-V designs and OpenTitan extensions for Kyber, Dilithium, and Falcon \cite{stelzer2025opentitan, karl2024hardware}. FPGA-based acceleration has also been shown to significantly speed up lattice-based operations \cite{he2023fpga}. However, while acceleration reduces computation energy, it does not mitigate the communication overhead of large post-quantum artifacts. Airtime, fragmentation, and packet overheads over PANs remain unchanged, which is precisely the regime examined in this study.

\section{Quantum Threats and Security Countermeasures} \label{QThreatsModel}

We consider an enhanced Dolev--Yao adversary with full control of the wireless channel, including eavesdropping, injection, modification, and replay capabilities \cite{dolevyao1983}. In addition, the adversary possesses future offline quantum computational capabilities and can retrospectively break classical public-key cryptography after observing protocol transcripts. This corresponds to a Harvest-Now--Decrypt-Later (HNDL) threat model. However, real-time quantum cryptanalysis during an active session is assumed impractical due to the expected latency and resource overheads of near-term quantum computing.

Under this model, the urgent requirement is a quantum-safe key establishment mechanism capable of deriving fresh session keys resistant to retrospective compromise. Once a session key is established, payloads protected using a 256-bit symmetric AEAD cipher maintain an adequate post-quantum security margin, since Grover's algorithm provides only a quadratic reduction in brute-force complexity \cite{grover1996search, grassl2016applying}.

We therefore realize a PQKE procedure directly over a PAN communication stack. As a representative implementation, we deploy the protocol over BLE. The design is intentionally minimal and backward-compatible \cite{liu2024towards}: it introduces a dedicated GATT service for key exchange over the ATT layer while leaving the remainder of the protocol stack unchanged. Although implemented on BLE, the core challenge---transporting large KEM artifacts over constrained link-layer frames---is common across PAN technologies.

We adopt a one-round KEM-based construction using ML-KEM due to its standardized status and balanced performance-security trade-off. We evaluate ML-KEM-512, ML-KEM-768, and ML-KEM-1024, corresponding to NIST security levels 1, 3, and 5 \cite{NIST_IR_8547_ipd}. The protocol proceeds as follows:

\begin{enumerate}
    \item \textbf{Key generation (Peripheral).}
    The peripheral generates an ML-KEM key pair $(pk, sk)$ and transmits the public key using \texttt{BLE Notify} operations.

    \item \textbf{Encapsulation and ciphertext transfer (Central).}
    The central encapsulates a shared secret using $pk$, producing ciphertext $ct$, and transmits $ct$ via \texttt{BLE Write} operations.

    \item \textbf{Decapsulation (Peripheral).}
    The peripheral decapsulates $ct$ using $sk$ to recover the shared secret and derives the session key.
\end{enumerate}

The devices subsequently protect application payloads using symmetric authenticated encryption under the derived session key.

\section{Energy Modeling and Measurement} \label{results}

In this section, we introduce a simple energy model to decompose those costs, and validate the resulting computation–communication contrast using measurements collected on real hardware.

\subsection{Theoretical Modeling}

We use an analytical model that decomposes total energy consumption into cryptographic computation and protocol-level communication components. The dominant contributors to these components correspond to four phases: key generation, public key transmission, ciphertext reception, and decapsulation. Accordingly, PQKE energy is approximated as the sum of these dominant operations:  

\begin{equation*}
E_{\mathrm{pqke}} = E_{\mathrm{comp}} + E_{\mathrm{comm}}  \approx E_{\mathrm{keygen}} + E_{\mathrm{Decap}} + E_{\mathrm{tx,pk}} + E_{\mathrm{rx,ct}}
\end{equation*}

Here, the computation energy terms ($E_{\mathrm{keygen}}$, $E_{\mathrm{Decap}})$ are estimated from the execution time of cryptographic primitives, obtained via cycle counts $C$ from the pqm4 project \cite{pqm4} and converted to energy using the MCU’s operating frequency $f$, as shown in Eq.~\ref{eq:Ecomp-cycles}. The communication energy terms ($E_{\mathrm{tx,pk}}, E_{\mathrm{rx,ct}}$) are estimated from the radio’s Tx and Rx current, the BLE PHY-layer data rate $R$, and the total PHY-layer bytes $B$, as shown in Eq. ~\ref{eq:Ecomm-air-time}.

\begin{align}
E_{\mathrm{keygen/decap}}
  &= I_{\mathrm{mcu}}\,V\,\frac{C_{\mathrm{keygen/decap}}}{f_{\mathrm{mcu}}}.
\label{eq:Ecomp-cycles}
\end{align}

\begin{align}
E_{\mathrm{tx/rx}}
  &= I_{\mathrm{tx/rx}}\,V\,\frac{B_{\mathrm{phy}}}{R_{\mathrm{phy}}}.
\label{eq:Ecomm-air-time}
\end{align}

In addition, secondary energy overheads arising from platform- and protocol-specific effects (e.g., OS scheduling, BLE internal operations, and device memory access) are captured through empirical measurements but are excluded from the analytical model.

\subsection{Empirical Measurement}

\subsubsection{Experimental Setup}

We implemented PQKE over BLE on an nRF52840 development kit and measured its energy consumption using a Nordic Power Profiler Kit II (PPK2) at a 100~kHz sampling rate. Cryptographic operations use the ML-KEM implementation from PQClean\cite{pqclean}, integrated into a Zephyr RTOS application adapted from \cite{liu2024towards}. Experiments used the BLE 1~Mbps PHY with a 50~ms connection interval and were powered by a 3.0~V DC source. Each experiment is repeated 10 times per ML-KEM variant, and reported values are averaged to reduce run-to-run variation.

\subsubsection{Empirical Measurement of $E_{\mathrm{comp}}$} \label{comp_factors}

\begin{figure}
    \centering
    \includegraphics[width=\linewidth]{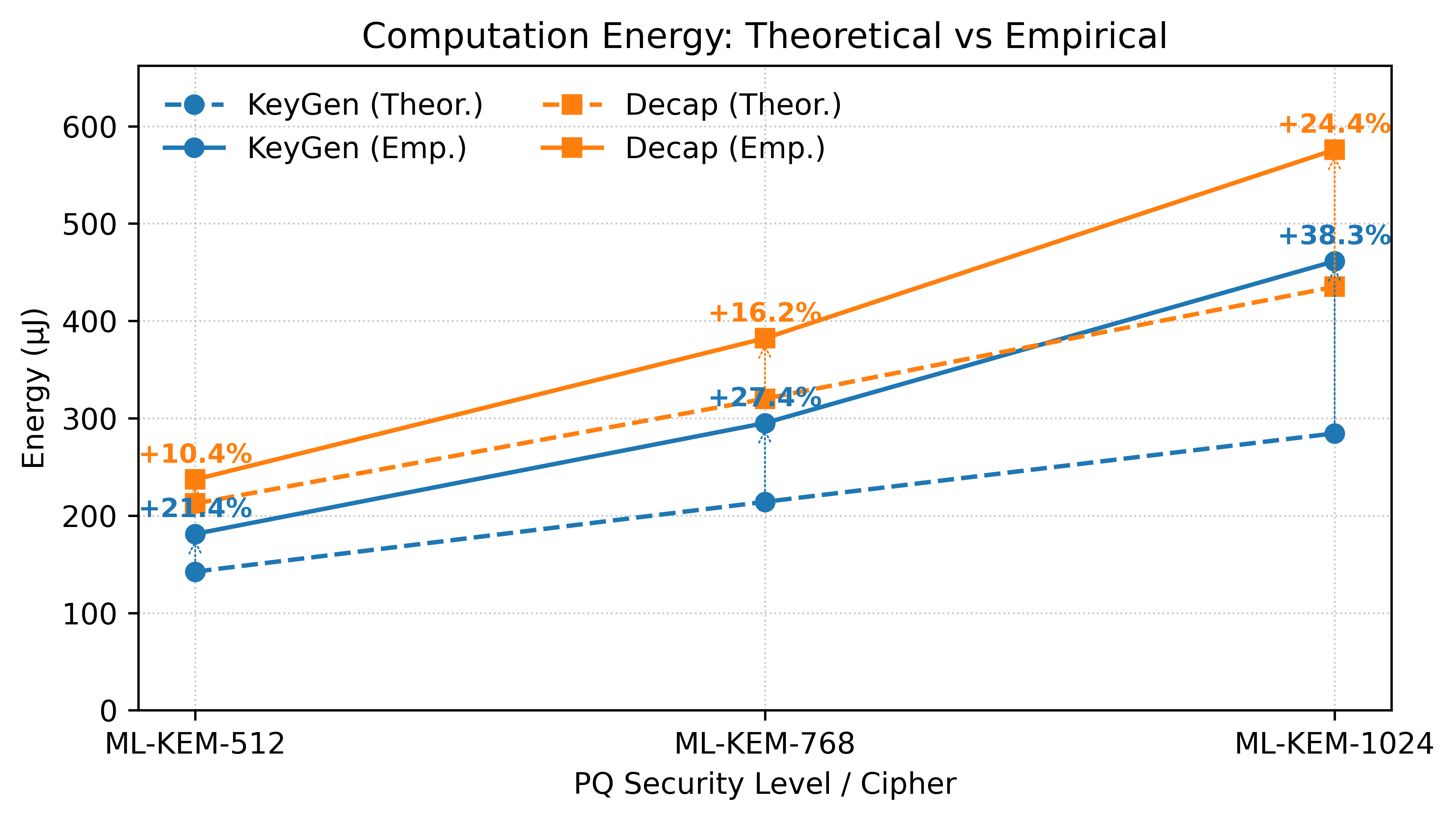}
    \caption{Theoretical (dashed) vs. empirical (solid) computation energy for ML-KEM key generation and decapsulation. +\% labels indicate the fraction of secondary cost over empirical.}
    \label{fig:comp_energy}
\end{figure}

We measured $E_{\mathrm{keygen}}$ and $E_{\mathrm{decap}}$ from PPK2 traces. The corresponding theoretical costs follow from the cycle-based model in Eq.~\eqref{eq:Ecomp-cycles}. We estimated the secondary cost for \texttt{KeyGen} and \texttt{Decap} as the residual between the measured and the estimated energy, and report this secondary cost as a share of the empirical total ($\Delta$). The results are visualized in Fig.~\ref{fig:comp_energy}, which shows that computation energy increases with PQ security level, and that the gap between measured energy and a cycles-only model widens accordingly. This underestimation is operation-dependent: \texttt{KeyGen} consistently exhibits a higher secondary share than \texttt{Decap}. This behavior is consistent with the differing system-level workloads of the two operations, with \texttt{KeyGen} involving more data handling and randomness generation, while \texttt{Decap} remains more compute-centric and tracks the cycle model more closely.

We then introduce a unified correction mechanism for computation energy. We express the adjusted computation energy as $E_{\mathrm{comp}}^{\mathrm{Adj}} = \gamma_\mathrm{comp}, E_{\mathrm{comp}}^{\mathrm{Theor}}$, where $\gamma_\mathrm{comp}$ is an empirically derived calibration factor. On our platform, calibration yields $\gamma_{\texttt{KeyGen}} \approx 1.27$, $1.38$, and $1.62$ for ML-KEM-512, ML-KEM-768, and ML-KEM-1024, respectively, while the corresponding $\gamma_{\texttt{Decap}}$ factors are approximately $1.12$, $1.19$, and $1.32$. 

\subsubsection{Empirical Measurement of $E_{\mathrm{comm}}$}

We further measured the energy associated with transmitting the public key (\texttt{Notify\_PK}) and receiving the ciphertext (\texttt{Write\_CT}). On the empirical side, we use the PPK2 trace and locate the CEs containing the corresponding ATT operations, then take each CE’s start and end from the first and last radio bursts in the trace. The operation’s empirical energy cost is the sum of these per-CE results. For each configuration we repeat the measurement across multiple runs and report the mean energy. 

On the analytical side, the communication cost is estimated from the total number of PHY-layer bytes exchanged during each operation, determined by ATT, L2CAP, and LL fragmentation rules. Each ATT fragment incurs a 3~B header, is encapsulated in a 4~B L2CAP header, and segmented into LL Data PDUs with a 10~B link-layer overhead per PDU; every LL transmission is acknowledged by an empty LL ACK and separated by a 150~µs inter-frame spacing (IFS). The total communication energy is computed by converting the resulting Tx and Rx PHY-layer bytes into on-air time using the data rate, adding the cumulative IFS duration, and multiplying the active times by the corresponding datasheet Tx, Rx, and IFS currents and the supply voltage, i.e., $E = IVt$.

\begin{table}[t]
\centering
\captionsetup[subtable]{justification=centering}

\begin{subtable}{\linewidth}
\centering
\caption{ML-KEM-512 / PQ Security L1}
\label{tab:comm_energy_512_ieee}
\resizebox{\linewidth}{!}{%
\begin{tabular}{cc ccc ccc}
\toprule
\multicolumn{2}{c}{\textbf{Pkt. Size (B)}} &
\multicolumn{3}{c}{\textbf{$E_{\mathrm{Notify\_PK}}$}} &
\multicolumn{3}{c}{\textbf{$E_{\mathrm{Write\_CT}}$}} \\
\cmidrule(lr){1-2}\cmidrule(lr){3-5}\cmidrule(lr){6-8}
\textbf{ATT} & \textbf{LL} &
\textbf{Theor. ($\mu$J)} & \textbf{Emp. ($\mu$J)} & \textbf{$\Delta$(\%)} &
\textbf{Theor. ($\mu$J)} & \textbf{Emp. ($\mu$J)} & \textbf{$\Delta$ (\%)} \\
\midrule
65  & 27  & 362.63 & 396.67 & 8.58\%  & 340.57 & 392.32 & 13.19\% \\
65  & 69  & 215.83 & 259.50 & 16.83\% & 200.68 & 217.10 & 7.56\%  \\
104 & 27  & 311.01 & 326.74 & 4.82\%  & 296.14 & 331.82 & 10.75\% \\
104 & 108 & 175.50 & 205.77 & 14.71\% & 167.44 & 190.97 & 12.32\% \\
204 & 27  & 306.71 & 308.89 & 0.71\%  & 291.94 & 326.55 & 10.60\% \\
204 & 208 & 148.61 & 170.56 & 12.87\% & 140.85 & 150.85 & 6.63\%  \\
404 & 27  & 304.56 & 313.89 & 2.97\%  & 284.24 & 325.98 & 12.81\% \\
404 & 251 & 146.46 & 168.97 & 13.32\% & 138.74 & 148.81 & 6.76\%  \\
\bottomrule
\end{tabular}}
\end{subtable}

\vspace{0.5\baselineskip}

\begin{subtable}{\linewidth}
\centering
\caption{ML-KEM-768 / PQ Security L3}
\label{tab:comm_energy_768_ieee}
\resizebox{\linewidth}{!}{%
\begin{tabular}{cc ccc ccc}
\toprule
\multicolumn{2}{c}{\textbf{Pkt. Size (B)}} &
\multicolumn{3}{c}{\textbf{$E_{\mathrm{Notify\_PK}}$}} &
\multicolumn{3}{c}{\textbf{$E_{\mathrm{Write\_CT}}$}} \\
\cmidrule(lr){1-2}\cmidrule(lr){3-5}\cmidrule(lr){6-8}
\textbf{ATT} & \textbf{LL} &
\textbf{Theor. ($\mu$J)} & \textbf{Emp. ($\mu$J)} & \textbf{$\Delta$ (\%)} &
\textbf{Theor. ($\mu$J)} & \textbf{Emp. ($\mu$J)} & \textbf{$\Delta$ (\%)} \\
\midrule
65  & 27  & 535.34 & 569.67 & 6.03\%  & 477.84 & 537.34 & 11.07\% \\
65  & 69  & 315.14 & 374.25 & 15.79\% & 281.99 & 319.64 & 11.78\% \\
104 & 27  & 464.63 & 482.46 & 3.69\%  & 420.12 & 458.81 & 8.43\%  \\
104 & 108 & 261.37 & 302.49 & 13.59\% & 235.46 & 267.14 & 11.86\% \\
204 & 27  & 458.18 & 456.83 & -0.29\% & 414.86 & 452.34 & 8.29\%  \\
204 & 208 & 221.04 & 247.69 & 10.76\% & 202.22 & 212.81 & 4.97\%  \\
404 & 27  & 449.31 & 467.44 & 3.88\%  & 406.11 & 475.49 & 14.59\% \\
404 & 251 & 217.81 & 244.72 & 10.99\% & 199.06 & 215.10 & 7.46\%  \\
\bottomrule
\end{tabular}}
\end{subtable}

\vspace{0.5\baselineskip}

\begin{subtable}{\linewidth}
\centering
\caption{ML-KEM-1024 / PQ Security L5}
\label{tab:comm_energy_1024_ieee}
\resizebox{\linewidth}{!}{%
\begin{tabular}{cc ccc ccc}
\toprule
\multicolumn{2}{c}{\textbf{Pkt. Size (B)}} &
\multicolumn{3}{c}{\textbf{$E_{\mathrm{Notify\_PK}}$}} &
\multicolumn{3}{c}{\textbf{$E_{\mathrm{Write\_CT}}$}} \\
\cmidrule(lr){1-2}\cmidrule(lr){3-5}\cmidrule(lr){6-8}
\textbf{ATT} & \textbf{LL} &
\textbf{Theor. ($\mu$J)} & \textbf{Emp. ($\mu$J)} & \textbf{$\Delta$ (\%)} &
\textbf{Theor. ($\mu$J)} & \textbf{Emp. ($\mu$J)} & \textbf{$\Delta$ (\%)} \\
\midrule
65  & 27  & 702.41 & 769.91 & 8.77\%  & 687.08 & 809.59 & 15.13\% \\
65  & 69  & 414.45 & 519.11 & 20.16\% & 407.29 & 475.54 & 14.35\% \\
104 & 27  & 612.61 & 638.42 & 4.04\%  & 603.81 & 661.99 & 8.79\%  \\
104 & 108 & 347.24 & 447.15 & 22.34\% & 340.81 & 345.97 & 1.49\%  \\
204 & 27  & 604.01 & 628.12 & 3.84\%  & 595.40 & 683.44 & 12.88\% \\
204 & 208 & 293.46 & 323.75 & 9.35\%  & 287.63 & 302.88 & 5.04\%  \\
404 & 27  & 594.06 & 590.63 & -0.58\% & 585.60 & 661.13 & 11.43\% \\
404 & 251 & 289.16 & 328.51 & 11.98\% & 283.42 & 302.56 & 6.33\%  \\
\bottomrule
\end{tabular}}
\end{subtable}

\caption{BLE communication energy breakdown across ATT MTU and LL PDU configurations for ML-KEM parameter sets. $\Delta$ denotes the secondary cost ratio, i.e., the residual overhead normalized to empirical energy.}
\label{tab:comm_energy_all}

\end{table}

Table~\ref{tab:comm_energy_all} compares theoretical and empirical communication energy. Empirical measurements consistently exceed theoretical estimates, indicating a moderate secondary energy component that accounts for approximately 5--20\% of the total communication energy. Fragmentation effects follow expected trends: increasing ATT MTU and LL PDU sizes reduces both theoretical and empirical energy by lowering protocol overhead. The Tx--Rx--IFS formulation captures these trends accurately, confirming airtime as the dominant contributor to BLE communication energy.

The residual gap between empirical and theoretical values is systematic across configurations and reflects hardware and firmware overheads not captured by the airtime-based model. To reconcile theory and measurement, we introduce a similar multiplicative correction: $E_{\mathrm{comm}}^{\mathrm{Adj}} = \gamma_{\mathrm{comm}} E_{\mathrm{comm}}^{\mathrm{Theor}}$. Across all tested BLE settings, $\gamma_{\mathrm{comm,\,cfg}}$ generally lies in the range of 1.10--1.20, with a nominal value of 1.15 providing a conservative yet accurate alignment between theory and measurement. 

\subsection{PQKE Energy Cost Breakdown}

With empirically calibrated computation and communication components, the total PQKE energy is expressed as:

\begin{align}
E_{\text{pqke}} &= E_{\text{comp}} + E_{\text{comm}} \notag\\
&\approx \gamma_{\text{KeyGen}} E_{\text{KeyGen}}^{\text{Theor.}}
 + \gamma_{\text{Decap}} E_{\text{Decap}}^{\text{Theor.}} \notag\\
&\quad + \gamma_{\text{comm}}\big(E_{\text{Notify\_PK}} + E_{\text{Write\_CT}}\big)
\label{Epqke_with_adj}
\end{align}

\begin{figure*}[h]
    \centering
    \includegraphics[width=\linewidth]{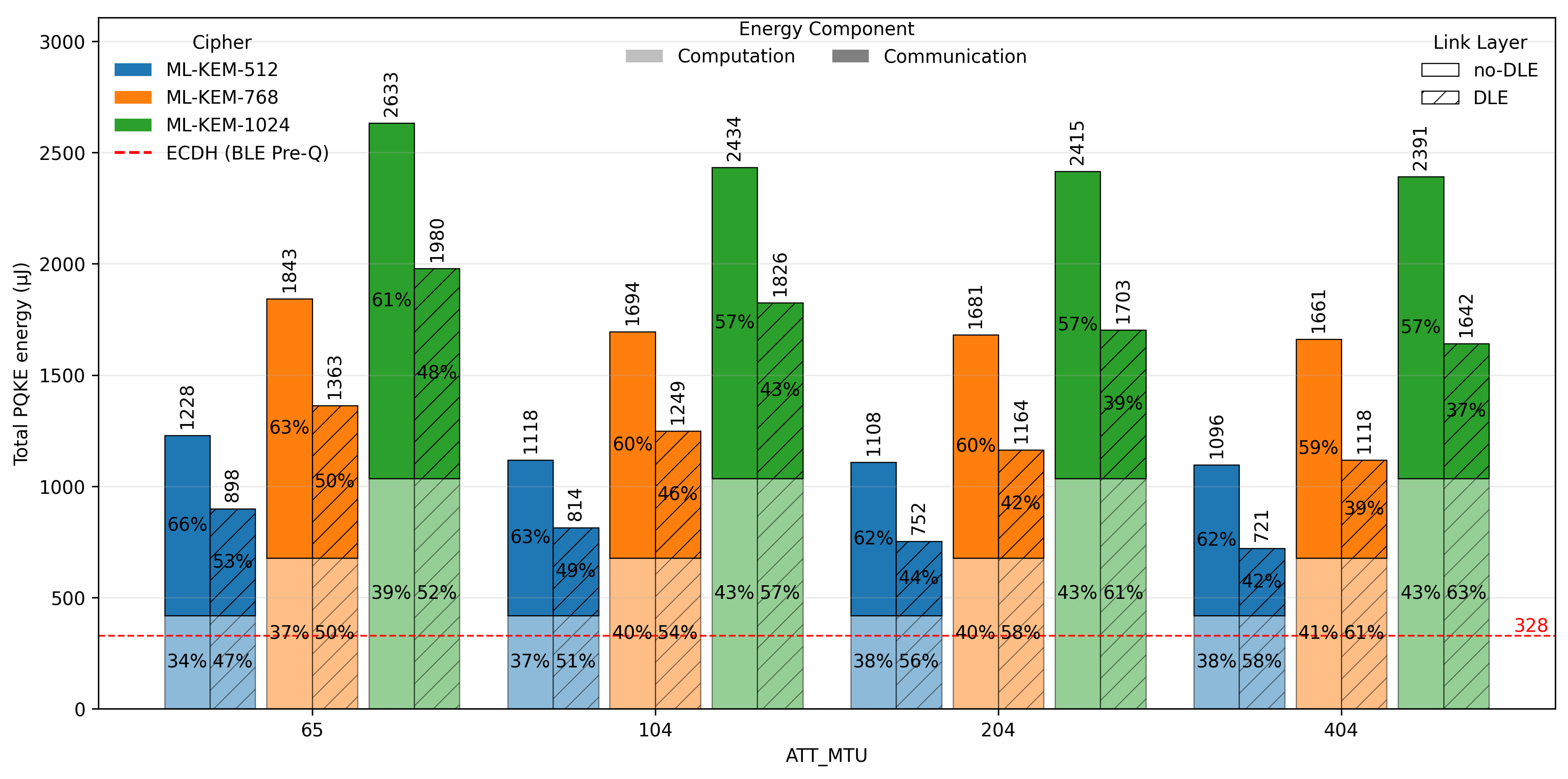}
    \caption{PQKE energy cost across BLE configurations, with computation–communication breakdown and DLE comparison.}
    \label{fig:pqke_stacked_all}
\end{figure*}

\noindent
Using this model, the resulting PQKE energy breakdown across ML-KEM variants and BLE configurations is shown in Fig.~\ref{fig:pqke_stacked_all}. For each ATT\_MTU setting, results are reported both under default BLE operation (LL PDU = 27) and with DLE enabled (large LL PDU which eliminates link-layer fragmentation), allowing direct comparison of fragmentation behavior. For reference, we also measured the cost of the default ECC pairing based on the NIST P-256 curve ($E_\text{ECDH}\approx328\mu J$). 

Across all configurations, the absolute PQKE energy ranges between 721--2633~$\mu J$, depending on security level and connection settings. Increasing the ML-KEM security level increases the computational component of PQKE, while variations in ATT\_MTU and LL PDU size primarily affect the communication component through changes in the number of link-layer fragments required to transmit post-quantum artifacts. Enabling DLE consistently reduces the total PQKE energy by lowering the communication overhead associated with fragmentation. The relative contribution of computation and communication varies with both security level and link-layer configuration, as indicated by the stacked composition and annotated percentages in Fig. \ref{fig:pqke_stacked_all}.

\subsection{Session-Level Energy Illustration}

\begin{figure}
    \centering
    \includegraphics[width=\linewidth]{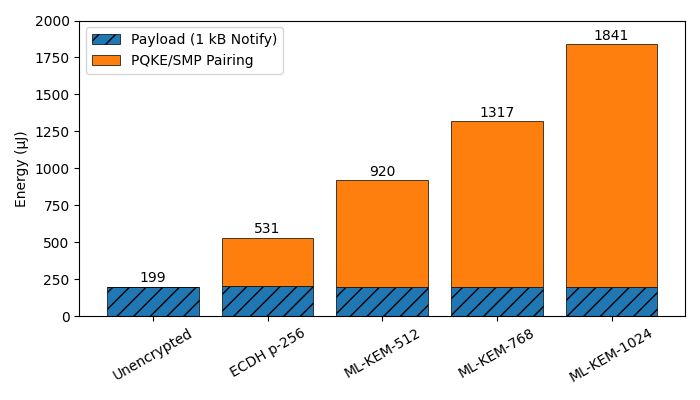}
       \caption{Total session energy for transmitting a 1~kB payload under different classic and quantum-safe security configurations.}
    \label{fig:3comp}
\end{figure}

To contextualize PQKE overhead in an application setting, we estimate the total energy required to establish a secure session and transmit a 1~kB payload using \texttt{BLE Notify} under the following configurations: (i) no security, (ii) classical pairing based on ECDH P-256 from Bluetooth SMP, and (iii) ML-KEM-512, ML-KEM-768, and ML-KEM-1024. All cases use optimized configuration with DLE enabled. The total session energy is computed as the sum of ECDH pairing/PQKE energy (from Section~4.3) and payload transmission energy using the communication model from Section~4.2. As shown in Fig.~\ref{fig:3comp}, for a single short message, pairing dominates total session energy, with PQKE incurring several times the cost of ECDH pairing and increasing systematically across PQ security levels.

\section{Discussion and Design Insights} \label{discussion}

The results reveal consistent structural patterns in PQKE on low-power personal-area networks. The total energy cost is shaped by two orthogonal factors: cryptographic complexity, which scales with the security level of the selected KEM, and protocol configuration, which determines how post-quantum artifacts are conveyed over a constrained wireless link. Compared to the existing BLE SMP pairing procedure, PQKE incurs an absolute energy cost approximately 2.5× to 8.5× higher. Importantly, this overhead is highly configuration-dependent: communication dominates in poorly optimized link-layer settings, while cryptographic computation becomes the primary contributor once fragmentation is reduced. This interplay reframes quantum-resilient pairing as a cross-layer optimization problem spanning cryptography, protocol design, and radio configuration, rather than a purely algorithmic concern.

\subsection{Energy Trade-Off Space: Computation vs Communication}
The energy composition of PQKE reflects a dynamic trade-off between computation and communication, jointly shaped by the chosen KEM parameters and the link-layer configuration. Computation energy, dominated by KeyGen and Decap, scales nearly linearly with the polynomial dimension of ML-KEM; as shown in Fig. \ref{fig:pqke_stacked_all}, moving from ML-KEM-512 to ML-KEM-1024 approximately doubles this cost. Communication energy, by contrast, is largely independent of the cryptographic parameter set and is instead determined by the fragmentation profile, i.e., the number of link-layer packets required to transport a cryptographic artifact.

This interaction produces a clear pivot in the PQKE energy landscape. At small LL PDUs, communication dominates, accounting for 57\%-- 63\% of total energy. As LL PDU and ATT MTU sizes increase, fragmentation overhead decreases sharply, reducing the communication share and exposing the computational floor. Once fragmentation is eliminated, computation becomes increasingly dominant, particularly for stronger ML-KEM variants (e.g., only 37\% communication share at ML-KEM-1024 with DLE). Overall, enabling DLE improves link-layer efficiency and reduces total PQKE energy by approximately 25–34\% across all security levels, highlighting the critical role of packet-level optimization in post-quantum protocol design.

\subsection{Quantum-Safe BLE Design Implications}

Although the experimental PQKE realization in this study was implemented as a custom BLE service, its link-layer behavior is governed by the same mechanisms that underpin standard BLE pairing. Consequently, the measured energy trends provide a realistic indication of how a future PQ-enabled BLE pairing procedure would behave under comparable link configurations. Viewed through this lens, our results highlight two key design implications for integrating post-quantum security into BLE systems.

\subsubsection{Minimize PQKE Energy via Link-Layer Optimization}

The results show that PQKE energy is often dominated by communication overhead, with link-layer fragmentation—rather than cryptographic computation—being the primary contributor. Enabling DLE substantially reduce the number of link-layer fragments required to carry post-quantum public keys and ciphertexts, thereby lowering cumulative header, acknowledgment, and inter-frame spacing costs. As such, PQKE should be treated as a short-lived, high-throughput transfer phase, during which link-layer parameters are temporarily optimized to minimize fragmentation. Importantly, this optimization can be confined to the pairing phase and does not interfere with the low-duty-cycle behavior that characterizes steady-state BLE operation.

\subsubsection{Balancing Security Strength and Energy Budget}

Beyond link-layer configuration, the measurements reveal a systematic trade-off between cryptographic security level and energy consumption. Higher security levels increase PQKE energy primarily through increased computational cost, while communication energy remains largely determined by BLE link settings and payload fragmentation. As a result, quantum-safe BLE designs should consider PQKE parameter selection as an explicit security–energy trade-off.

Importantly, our calibrated empirical model provides quantitative guidance for this decision: by separating computation and communication components and incorporating the correction factors $\gamma_{\mathrm{comp}}$ and $\gamma_{\mathrm{comm}}$, designers can directly estimate the incremental energy cost of moving from ML-KEM-512 to higher security levels under specific BLE configurations. Energy-constrained or energy-harvesting devices may preferentially adopt lower-security variants to maintain feasibility, whereas devices with larger energy budgets can support stronger post-quantum security levels when required. This flexibility enables PQ-enabled BLE systems to adapt security strength to device capability without fundamentally altering the underlying protocol stack.

\subsection{Toward Quantum-Resilient PANs Beyond BLE}

While this study focused on connection-oriented BLE as a representative platform, the findings reveal a broader pattern applicable to other PANs. Across technologies, the introduction of PQC amplifies the fundamental tension between cryptographic complexity and communication efficiency. In BLE, PQKE already incurs substantial communication energy due to fragmentation and link-layer timing, despite tightly scheduled CEs with direct acknowledgments. In contention- and flooding-based protocols such as BLE Mesh and IEEE 802.15.4, probabilistic medium access, retransmissions, and redundant relaying further extend radio activity, exacerbating this imbalance. As a result, if PQKE energy is communication-dominated even in connection-oriented BLE, it is likely to become an even more pronounced network-level energy bottleneck in contention-based PANs.

To maintain energy feasibility under post-quantum payload sizes, several cross-layer design principles emerge and may be investigated in future works:

\subsubsection{Schedule or bound contention during PQKE phases}
When large cryptographic artifacts must be exchanged, normal background traffic should be temporarily reduced to avoid collisions. Gateways or coordinators can allocate short “secured access windows” for PQKE traffic, effectively restoring the predictability lost in contention-based access and limiting the exponential growth of retry events.

\subsubsection{Limit redundancy through adaptive reliability policies}
Both CSMA/CA retries (IEEE 802.15.4) and BLE Mesh flooding rely on probabilistic repetition. Introducing adaptive policies that adjust retry counts, relay density, or transmission intervals according to channel conditions can preserve reliability without wasting energy. 

\subsubsection{Align fragmentation strategy with contention behavior}
Because contention costs grow with the number of transmission attempts, reducing the fragment count of PQ artifacts directly curtails energy. However, in highly loaded networks, longer fragments may suffer higher collision probability. A protocol-aware fragmentation scheme may strike a balance between transmission duration and retry likelihood.

\subsubsection{Exploit role asymmetry in network architecture}
Many PANs have clear energy and capability hierarchies between gateways and end devices. Offloading PQ computation and communication to better-powered nodes can redistribute the energy burden. 

\section{Conclusion} \label{conclusion}
This study analyzed the energy cost of post-quantum key establishment in low-power PANs using BLE as a representative platform. We decomposed PQKE into computation and communication components and validated this model through empirical measurements on real hardware. The results show that PQKE energy is strongly configuration-dependent: while unoptimized communication can dominate cost and exceed that of classical BLE pairing, link-layer optimization significantly reduces overhead. Feasibility is therefore governed less by cryptographic arithmetic than by how efficiently large cryptographic artifacts are transported over the wireless link. Although demonstrated on connection-oriented BLE, these findings extend to low-power PANs more broadly, particularly to contention-based protocols where retransmissions and channel access overheads can further amplify communication cost. 

Overall, our results demonstrate that quantum resilience is practical on constrained devices when pairing is treated as a cross-layer design problem that jointly considers cryptography and link-layer efficiency.

\balance

\bibliographystyle{ACM-Reference-Format}
\bibliography{bibtex/bib/ref}

\end{document}